Title:   A Model of Late Long-Term Potentiation Simulates Aspects of Memory Maintenance

Author:  Paul Smolen

Laboratory of Origin:

Department of Neurobiology and Anatomy
W.M. Keck Center for the Neurobiology of Learning and Memory
The University of Texas Medical School at Houston
P.O. Box 20708
Houston, TX 77225

Correspondence Address:

Paul D. Smolen
Department of Neurobiology and Anatomy
W.M. Keck Center for the Neurobiology of Learning and Memory
The University of Texas-Houston Medical School
P.O. Box 20708
Houston, TX 77225
Voice: (713) 500-5564
FAX: (713) 500-0623
E-mail: Paul.D.Smolen@uth.tmc.edu

# INTRODUCTION

Considerable evidence supports the hypothesis that long-term potentiation (LTP) of synaptic connections is an essential component of memory formation and maintenance [1,2,3]. Indeed, reversal of LTP maintenance correlates with loss of spatial memory [4]. This hypothesis implies specific memories are at least partly represented by patterns of strengthened synapses. Neurons connected by these synapses would be more likely to fire in a correlated manner. An outstanding question of neurobiology is, how can some memories be preserved for months or years given turnover of neuronal and synaptic proteins? Synaptic proteins typically have lifetimes of hours to a few days [5]. Furthermore, synaptic homeostasis mechanisms that act to normalize neuronal activity or excitatory synaptic drive have been identified [6,7,8] and might act to return potentiated synapses to an average or basal strength, eliminating memories.

To preserve strong synapses despite these processes, some proposed mechanisms hypothesize biochemical positive feedback loops. Synaptic strengthening is assumed to accelerate a process such as translation of synaptic proteins, which itself drives synaptic strengthening. Bistability in synaptic weights could be generated by such a feedback loop. Some specific examples are, 1) A postulated feedback loop based on reciprocal activation of protein kinase C, MAP kinase (MAPK), and phospholipase A2 [9,10]; 2) A loop in which transient enhancement of translation of an elongation factor, eEF1A, leads to self-reinforcing translation of eEF1A and other mRNAs required to consolidate L-LTP [11,12]; and 3) A loop in which transient activation of protein kinase A (PKA) phosphorylates a critical number of AMPA receptors and saturates the activity of a receptor phosphatase, after which basal PKA activity suffices to maintain phosphorylation of these receptors [13].

In models incorporating positive feedback and bistability, potentiation switches synapses to the upper weight state. These models represent candidate mechanisms for long-term memory storage [14,15]. However, these models have not yet suggested mechanisms for selective preservation of particularly important or relevant memories. Also, Fusi et al. [16] has argued that bistability based on biochemical positive feedback is unlikely to suffice for biological memory storage. The expected lifetime of memory with this type of bistability scales only as the logarithm of the number of synapses storing the memory. For physiologically reasonable rates of plasticity-inducing events, memories stored by even a large number of synapses ($> 10^6$) would be likely to decay in < 1 hr. The authors suggest one solution – the expansion of each synaptic weight state into a cascade of states connected by metaplastic biochemical transitions. I suggest that even without such cascades of states, bistability based on reinforcement of strengthened synapses by ongoing, "background" neuronal activity could sustain biological memory storage.

The concept that reactivation of modified synapses may be necessary for long-term memory maintenance has a history of several decades [17,18]. However, experimental support has been obtained only recently. Shimizu et al. [19] used an inducible knockout of the NMDA receptor (NMDAR) in hippocampal region CA1 to demonstrate that continued NMDAR function after training is necessary for maintenance of contextual fear conditioning over 1 month. Cui et al. [20] showed that inducing NMDAR knockout 6 months after formation of contextual or cued fear memories disrupted retention of these remote memories. The retention locus is not hippocampal (at least for cued memories) and is likely neocortical. Furthermore, knocking out

NMDARs disrupts cortical storage of nondeclarative taste memory [21]. Some authors have hypothesized that reactivation of synaptic circuits encoding memory traces occurs largely during sleep [17]. Consistent with this, Qin et al. [22] found that correlations between hippocampal and cortical neuron activities, established during learning, could be measured during sleep.

Some previous models have suggested reactivation may preferentially occur at strong synapses. Such preferential reactivation has been termed synaptic reentry reinforcement (SRR) [23,24]. Preservation of patterns of strengthened synapses has been simulated [24,25,26], suggesting that SRR can allow patterns of strong synapses, and associated memories, to be maintained for the life of an animal. One investigation illustrates how bistability of synaptic weights can arise from a positive feedback loop between excitatory synaptic weight increase and ongoing neuronal activity [27], which is similar to SRR.

However, models that include SRR have <u>not</u>, so far, incorporated representations of neuronal biochemical events necessary for late long-term potentiation (L-LTP, lasting for > 2-3 hrs). These events include $Ca^{2+}$ influx, kinase activation, and gene induction. Induction of hippocampal L-LTP by tetanic stimuli is accompanied by upregulation of numerous, clustered genes [28]. Transcription is required for late L-LTP [29] although translation suffices for the first 4-6 hrs after L-LTP induction [30]. Stimuli that induce L-LTP also set a synaptic "tag" that enables recently active synapses to selectively "capture" newly synthesized proteins [31,32]. It is likely that SRR depends at least partly on repeated reactivation of the biochemical pathways involved in L-LTP. SRR thus appears to be based on positive feedback in which stronger synapses are more frequently reactivated, resulting in repeated occurrences of L-LTP. To relate SRR to L-LTP, I formulated a reduced version of a previous biochemical model [33] that described L-LTP induction by different stimulus protocols. The reduced model represents in a more simplified and intuitive way the biochemical nonlinearities that transduce activity-induced $Ca^{2+}$ influx into synaptic tagging and gene induction.

The reduced model simulated L-LTP induction by tetanic or theta-burst protocols. It also simulated L-LTP induced by the original protocol demonstrating synaptic tagging [32]. Memory maintenance was then simulated at a single synapse or in a small group of synapses. Simulations first examined maintenance of the simplest possible "memory" – strengthening of one synapse. Ongoing neuronal activity was simulated by imposing continuous brief elevations of synaptic $Ca^{2+}$ at a frequency corresponding to theta oscillations. The reduced model represents a positive feedback loop between ongoing neuronal activity and synaptic weight increase, suggesting a mechanism for SRR-based long-term memory maintenance. Bistability of synaptic weight was generated by this feedback. The weight values in the stable states were reasonably robust to modest variations in model parameters. A synapse could be switched to the high weight state by a simulated tetanus or theta-burst, and subsequently remained strong despite minor fluctuations in background synaptic activity. Thus, this simple memory could be stored indefinitely.

In the SRR hypothesis groups of strengthened synapses, corresponding to specific memories, are episodically reactivated, maintaining the strength of those synapses. Forgetting will occur if a pattern is not reactivated for a long time. The experiments in which NMDAR knockout disrupted maintenance of remote memories [20,21] are compatible with the hypothesis that prolonged blockage of reactivation cortical synapses results in memory decay and loss. The normal episodic reactivation needed for memory maintenance could occur during sleep as noted above, and during conscious recall. Because reactivation is episodic, a model of memory maintenance via SRR should simulate synaptic weight maintenance by episodic ongoing activity.

These dynamics were observed with the reduced model. Bistability, and permanent synaptic strengthening by tetani, were seen when ongoing activity consisted of brief periods of $Ca^{2+}$ elevation separated by intervals (timescale of minutes) without activity. A high synaptic weight state, once formed, could be maintained with longer inactive intervals (timescale of ~ 1 day). However, prolonged cessation of synaptic activity would return a synapse to the low weight state. Subsequently, L-LTP could again occur. This mechanism for memory loss was simulated in a small group of synapses. L-LTP was repeatedly induced at all synapses, and periodic prolonged decreases in ongoing activity were imposed at all but one of the synapses. At the synapses with activity decreases, repeated cycles of potentiation and weakening occurred. The synapse with constant high activity retained its memory. It is plausible that this cyclic mechanism for controlling bistability of synaptic weight may correspond to a mechanism for biological memory forgetting, making synapses available for formation of new memories. Synapses that encode older memories might be less frequently reactivated, leading to selective loss of such memories.

Ongoing synaptic activity was also incorporated into the previous, detailed model of L-LTP induction [33]. Bistability of synaptic weight was obtained, and is similarly sustained by positive feedback between synaptic activity and synaptic weight. The bistability was robust to minor parameter variations. Tetanic stimuli could switch the synaptic weight to the upper state. Thus, both models simulate memory maintenance based on reactivation of biochemical pathways involved in L-LTP. These bistable models of memory maintenance generate experimental predictions (see Discussion). Tests of these predictions will improve understanding of the coupling between memory maintenance and neuronal activity.

## METHODS

### Reduction of a detailed model of late long-term synaptic potentiation

The previous detailed model [33] represented activation of CaM kinases, MAPK, and protein kinase A (PKA) subsequent to electrical or chemical synaptic stimuli. $Ca^{2+}$ influx activated CaM kinases II and IV (CAMKII, CAMKIV). cAMP elevation activated PKA. Stimuli also activated the MAPK cascade, in which Ras kinase phosphorylates and activates MAP kinase kinase, which activates MAPK. The tag that "marks" synapses for L-LTP involves covalent modifications that allow a synapse to "capture" plasticity factors (proteins or mRNAs) and incorporate them to increase synaptic strength [31]. PKA appears necessary for at least one of these modifications [34]. Postsynaptic CAMKII activation is also required for tetanic L-LTP at CA3-CA1 synapses [35] and synaptic MAPK may contribute by phosphorylating regulators of translation [36]. Therefore, setting a synaptic tag was postulated to require CAMKII, MAPK, and PKA activation. The variable TAG represented the level of the tag. The model further postulated that stimuli elevated nuclear $Ca^{2+}$, activating CAMKIV and MAPK. Empirically, MAPK activation leads to phosphorylation of transcription factors such as CREB and Elk-1 [37] and gene clusters upregulated during tetanic L-LTP correlate with CREB binding sites [28]. CAMKIV phosphorylates the transcriptional cofactor CREB binding protein [38], and CAMKIV inhibition blocks L-LTP [39]. In the model, CAMKIV and MAPK phosphorylate two unspecified transcription factors, inducing a gene. The level of expressed protein was denoted GPROD. The rate of increase of a synaptic weight W was set equal to the product of GPROD and TAG. The increase in W is limited by the concentration [P] of an unspecified synaptic protein. For equations and parameters, see [33].

To reduce the detailed model, I first assumed stimulus-induced $Ca^{2+}$ elevation is the common signal initiating synaptic tagging and gene induction. Recent observations support this assumption. $Ca^{2+}$ elevation activates adenylyl cyclases, producing cAMP that is critical for PKA activation and hippocampal L-LTP [40]. Activation of the neuronal MAPK cascade may be largely downstream of $Ca^{2+}$ influx and CaM kinase I activation [41]. Neuronal depolarization increases nuclear CaM kinase I activity [42], which could activate nuclear MAPK. In the model, the two differential equations describing induction of GPROD and TAG were coupled directly to elevations in nuclear and synaptic $Ca^{2+}$ respectively. Representations of specific kinases were eliminated. Figure 1A schematizes this reduced model. This model and the previous detailed model were developed to represent electrically induced L-LTP in the hippocampal CA3-CA1 pathway. Most of the characterized biochemical events important for this L-LTP are postsynaptic. Therefore, that model and the reduced model both focus on postsynaptic events.

Steeply sigmoidal, supralinear functions of nuclear and synaptic $Ca^{2+}$, were used respectively in the terms giving the rates of increase of TAG and GPROD. These functions are denoted $Th_1$ and $Th_2$ in Fig. 1A. The reduced model has only three ordinary differential equations. These equations describe the dynamics of TAG, GPROD, and the synaptic weight W. The differential equation for TAG represents the rate of synaptic tagging as an extremely steep, "threshold" function of the level of synaptic $Ca^{2+}$, $Ca_{syn}$, as follows,

$$\frac{d(TAG)}{dt} = k_{phos} S_T^4 (1 - TAG) - k_{deph} TAG \qquad 1)$$

with

$$S_T = \frac{Ca_{syn}^4}{Ca_{syn}^4 + K_{ST}^4} \qquad 2)$$

The first term on the right-hand side of Eq. 1 expresses the rate of increase of TAG as proportional to the fourth power of a Hill function of the level of $Ca_{syn}$ ($S_T$, Eq. 2). This very steep, supralinear product of Hill functions represents the convergence of the multiple kinases required for synaptic tagging. The activity of each kinase is in turn a supralinear function of $Ca^{2+}$. The Hill function $S_T$, with Hill coefficient of 4, represents activation of CaM kinases or adenylyl cyclase by $Ca^{2+}$, because the initial step in activation is cooperative binding of four $Ca^{2+}$ ions to calmodulin (CaM) [43,44]. Activation of Ras, the initial kinase in the MAPK cascade, is also strongly supralinear with respect to $Ca^{2+}$ concentration (Hill coefficient ≥4, [45]). Thus, in Eq. 1, three powers of $S_T$ represent activation of CAMKII, $Ca^{2+}$–sensitive adenylyl cyclase, and Ras. The fourth power of $S_T$ in Eq. 1 could represent the requirement for phosphatidylinositol-3-kinase (PI3K) in L-LTP [46]. Activation of PI3K may require CaM-$Ca_4$, because PI3 kinases contain CaM target sequences, and in CHO cells can activate PI3K [47]. Alternatively, the fourth power of $S_T$ could represent increased synthesis of the constitutively active PKC isoform PKMζ, if that increase is $Ca^{2+}$ – dependent. Increased translation of PKMζ is necessary for induction and maintenance of L-LTP [48,4] and PKMζ acts in the vicinity of the synapse [49].

Eq. 1 represents an extremely steep relationship between the rate of TAG formation and the level of $Ca^{2+}$. However, similarly steep stimulus-response relationships with Hill coefficients in excess of 20 have been observed and modeled in other biochemical pathways (activation of MAPK by progesterone, [50]; activation of gene expression by NF-κB, [51]). The sensitivity of the model to a decrease in the power of $S_T$ in Eq. 1 is examined in Results.

The differential equation for GPROD is similar to Eq. 1. GPROD synthesis is a supralinear function of nuclear $Ca^{2+}$ concentration,

$$\frac{d(GPROD)}{dt} = k_{syn} S_P^3 - k_{deg} GPROD \qquad 3)$$

with

$$S_P = \frac{Ca_{nuc}^4}{Ca_{nuc}^4 + K_{SP}^4} \qquad 4)$$

The first term on the right-hand side of Eq. 3 contains a third power of a sigmoidal function $S_P$. $S_P$ is a Hill function of nuclear $Ca^{2+}$, $Ca_{nuc}$, with Hill coefficient of 4 to represent $Ca^{2+}$ binding to CaM. Two powers of $S_P$ represent activation of CAMKIV and MAPK by CaM-$Ca_4$. Active MAPK and CAMKIV induce GPROD. The third power of $S_P$ is used because for CAMKIV to be activated, CaM-$Ca_4$ must also bind to the upstream enzyme CAM kinase kinase.

In the reduced model, brief $Ca^{2+}$ elevations induce brief, sharp increases in TAG and GPROD. In the detailed model [33] intermediate processes between $Ca^{2+}$ elevation and GPROD elevation were represented (*e.g.*, transcription factor phosphorylation). This leads to a slower increase in GPROD, but does not significantly alter the time course of synaptic weight changes.

The rate of increase of the synaptic weight, W, can be assumed proportional to the product of TAG and GPROD. However, these dynamics imply, and simulations confirmed, that W would increase to unreasonable levels as the number of tetanic stimuli increased. In the previous model [33] increase in W was limited by assuming an additional protein, P, was required and used up by increases in W. For the reduced model, a simpler approach was used, avoiding additional equations. An upper bound for W was imposed. Because W is non-dimensional, the upper bound was scaled to 1. Also small basal rates of synaptic strengthening ($k_{ltpbas}$) and depression ($k_{ltdbas}$) were assumed to occur in the absence of synaptic tagging. The resulting differential equation is,

$$\frac{d(W)}{dt} = \left\{ k_{ltp}(TAG)(GPROD) + k_{ltpbas} \right\}(1 - W) - k_{ltdbas} W \qquad 5)$$

**Simulation of L-LTP – inducing stimuli**

Concurrent elevations of $Ca_{syn}$ and $Ca_{nuc}$ represent electrical stimuli in the reduced model. Recent data [52] indicate that $[Ca^{2+}]$ transients in the vicinity of synaptic contacts, in dendritic spines, have a high amplitude and rapid decay. An action potential gives a $Ca^{2+}$ elevation of about 0.5 μM, with a decay time constant of only ~20 msec. Considering these data, the $Ca_{syn}$ elevation due to a 1-sec tetanus was modeled as a 1-sec elevation to ~1 μM (the parameter $Ca_{tet}$, Table I). Experimental data do not constrain well the duration or amplitude of nuclear $Ca^{2+}$ elevation following a tetanic stimulus. For the reduced model, a 1-sec elevation to a level of ½×$Ca_{tet}$ was assumed. The longer $Ca^{2+}$ elevation due to a theta-burst stimulus was also simulated. In theta-burst stimuli, 10-12 bursts of four 100 Hz pulses are typically delivered over a duration of ~ 2.2 sec [53]. This protocol was simulated with a single 2.5-sec square-wave increase in $Ca_{syn}$ to $Ca_{tet}$, and $Ca_{nuc}$ to ½×$Ca_{tet}$. In the absence of stimuli, $Ca_{syn}$ and $Ca_{nuc}$ remained at a basal level, $Ca_{bas} = 0.04$ μM.

## Numerical methods

The forward Euler method was used for integration, with a time step of 36 msec (Figs. 2-6 and Fig. 8) or 1.5 msec (Fig. 7). I verified that time step reductions by factors of 2 or 4 did not significantly alter the results. Initial values for model variables were determined as follows. Prior to L-LTP induction, all model variables were equilibrated for 100 hrs of simulated time with synaptic and nuclear $Ca^{2+}$ fixed at the basal value, $Ca_{bas} = 0.04$ μM. During equilibration only, to ensure equilibration, the variables with the slowest time constants (W, GPROD, and [P]) were set equal to their steady-state values determined by setting their derivatives to zero. Variables are non-dimensional, except for $Ca^{2+}$ concentrations (μM). Programs are available upon request.

## Modeling memory maintenance

Simulations of memory maintenance examined strengthening of either a single synapse or a group of 5 synapses. For the latter case, the synapses converge onto a single neuron, which is represented by only two variables. These are the nuclear $Ca^{2+}$ concentration, $Ca_{nuc}$, and the concentration of induced gene product, GPROD. Therefore, in simulating the group of synapses, a common $Ca_{nuc}$ and a common GPROD are coupled identically to all the synapses. Fig. 1B schematizes this model variant. For each synapse in the group, Eqs. 1, 2, and 5 were duplicated. Indexing synapses by $i = 1, ..., N_{syn}$, the network variants of Eqs. 1, 2, and 5 are

$$\frac{d\left(TAG_i\right)}{dt} = k_{phos} S_{T,i}^4 \left(1 - TAG_i\right) - k_{deph} TAG_i \qquad 6)$$

$$S_{T,i} = \frac{Ca_{syn,i}^4}{Ca_{syn,i}^4 + K_{ST}^4} \qquad 7)$$

$$\frac{d\left(W_i\right)}{dt} = \left\{k_{ltp}(TAG_i)(GPROD) + k_{ltpbas}\right\}(1 - W_i) - k_{ltdbas} W_i \qquad 8)$$

To simulate memory maintenance, ongoing $Ca^{2+}$ transients at each synapse were assumed to represent background neuronal activity. These brief elevations in $Ca_{syn}$ occurred with a period of 180 ms, within the range of hippocampal theta rhythm. Between elevations $Ca_{syn} = Ca_{bas}$. The amplitudes of the $Ca_{syn}$ elevations were determined as follows. For each synapse in a group, or for a single synapse, the maximal $Ca_{syn}$ elevation was hypothesized to be a saturating function of the synaptic weight, represented by the following equations,

$$S_W = \frac{W}{W + K_{max}} \qquad\qquad 9)$$

$$Ca_{max} = S_W\, Ca_{limit} \qquad\qquad 10)$$

As W increases to its maximal value of 1, Eqs. 9 and 10 imply the maximal $Ca_{syn}$ elevation during ongoing activity increases to 0.71 μM (using $K_{max}$ and $Ca_{limit}$ values from Table I). This increase in $Ca^{2+}$ with synaptic weight is essential for simulation of synaptic strength maintenance. A synaptic weight increase generates higher synaptic $Ca^{2+}$ elevations, which in turn induces L-LTP, reinforcing the weight increase. This positive feedback loop can generate bistability of synaptic weight. The argument supporting this hypothetical increase in $Ca^{2+}$ with synaptic weight is given in the Discussion.

Ongoing synaptic $Ca^{2+}$ elevations are likely to induce nuclear $Ca^{2+}$ elevation and gene induction, contributing to reinforcement of synaptic weight. The elevation of $Ca_{nuc}$ above $Ca_{bas}$ was hypothesized to be a saturating function of the total synaptic activity. The sum over synapses of the maximal $Ca_{syn}$ elevation represented the total synaptic activity. These assumptions result in the following equations for $Ca_{nuc}$,

$$Ca_{sum} = Ca_{max,1} \quad \text{(simulations with 1 synapse)}, \qquad\qquad 11a)$$

$$Ca_{sum} = \sum_{i=1}^{Nsyn} Ca_{max,\,i} \quad \text{(simulations with a group of synapses)} \qquad\qquad 11b)$$

$$Ca_{nuc} = \frac{f_{nuc} Ca_{sum}}{Ca_{sum} + N_{syn} K_{sum}} + Ca_{bas} \qquad\qquad 12)$$

From Eqs. 9-12, $Ca_{nuc}$ is a saturating function of each synaptic weight $W_i$ and of their sum. Similar saturating relationships of neuronal activity (here represented by $Ca_{nuc}$) to the sum of synaptic weights or activities are commonly used to model neuronal networks (*e.g.*, [54]).

The ongoing $Ca^{2+}$ elevations are not expected to all have identical amplitude $Ca_{max}$. Rather, the variable nature of neuronal activity generates a distribution between 0 and $Ca_{max}$. I chose a uniform distribution over this range and did not consider any correlation between successive elevations. The simulation time step was 36 msec (0.00001 hr). To simulate the distribution of $Ca^{2+}$ elevations, a random number uniformly distributed between 0 and 1 was chosen for each synapse at every fifth time step (every 180 msec). The elevation of $Ca_{syn}$ lasted for one time step, during which $Ca_{syn}$ was given by the following equation,

$$Ca_{syn,\,i} = Ca_{bas} + rand_i \cdot \left( Ca_{max,\,i} - Ca_{bas} \right) \qquad\qquad 13)$$

The frequency of these $Ca^{2+}$ elevations (5.6 Hz) is similar to theta rhythm.

Repeated loss of synaptic strength due to reduction of ongoing activity, as well as repeated gain of strength due to LTP, was simulated with a group of 5 synapses (Fig. 8 in Results), using Eqs. 6-13. In order to simulate loss of strength, it was found necessary to incorporate a simple representation of a synaptic homeostasis mechanism, synaptic scaling. Empirically, activity-dependent synaptic scaling has been observed in hippocampal slice [6,8] and in cortical cultures and other preparations (reviewed in [7]). Reduction of neuronal activity and enhancement of activity elicits, respectively, increases *vs.* decreases in synaptic strength as measured by the amplitude of excitatory postsynaptic currents. For the reduced model, for Fig. 8 only, a simple phenomenological representation of scaling was incorporated. For all synapses in the group, the upper bound of the synaptic weight was no longer fixed at 1.0 as in Eq. 5. Instead the bound was given by a maximal value ($W_{max}$) minus the product of an increment ($W_{inc}$) with the number of strong synapses ($W > 0.4$, corresponding to the upper stable state of W). Therefore, Eq. 5 was modified as follows,

$$\frac{d(W)}{dt} = \left\{ k_{ltp}(TAG)(GPROD) + k_{ltpbas} \right\} \left[ \left( W_{max} - N_{(W>0.4)} W_{inc} \right) - W \right] - k_{ltdbas} W \quad 14$$

To ensure W remains positive in simulations, $W_{max}$ must be large enough to always exceed $N_{(W>0.4)} W_{inc}$. In simulations with Eq. 14, the weight of any given strong synapse was seen to decrease in a stepwise manner as other synapses were strengthened.

Standard values for all model parameters in Eqs. 1-14 are in Table 1 and were used unless otherwise indicated.

## Simulation of a synaptic tagging protocol

Induction of L-LTP by the protocol of [33] was simulated. Two synapses were assumed to converge onto one neuron. Equations 1, 2, and 5 were duplicated for each synapse. Eqs. 3-4 were used to describe GPROD dynamics. Ongoing synaptic activity was not present, so $Ca_{syn,1}$, $Ca_{syn,2}$, and $Ca_{nuc}$ equaled $Ca_{bas}$ in the absence of stimuli. Three 1-sec tetani were delivered to synapse 1, with 10-min intervals between tetani. Tetanus parameters were as described above. $Ca_{syn,1}$, $TAG_1$, $Ca_{nuc}$, GPROD, and $W_1$ were elevated as a result. In the experimental protocol, anisomycin was applied 35 min later. Anisomycin blocks protein synthesis, which corresponds to blocking GPROD synthesis in the model. Therefore, the GPROD synthesis rate constant $k_{syn}$ was set to zero 35 min after the synapse 1 tetani and remained zero thereafter. One hour after the synapse 1 tetani, three 1-sec tetani were delivered to synapse 2 at 10-min intervals.

## Simulation of memory maintenance with the detailed model of L-LTP induction

For the simulation of Fig. 7 in Results, the detailed model [33] was used to describe the effect of electrical stimuli on synaptic weight. L-LTP was induced by three tetani at 5-min intervals. In this model, tetani briefly and concurrently elevate four stimulus variables: $Ca_{syn}$, $Ca_{nuc}$, [cAMP], and a rate constant $k_{f,Raf}$ for activation of Raf kinase. Each tetanus was modeled by brief concurrent elevations of $Ca_{syn}$ to 1.0 μM, $Ca_{nuc}$ to 0.5 μM, [cAMP] to 0.15 μM and $k_{f,Raf}$ to 0.24 min$^{-1}$. $Ca^{2+}$ elevations lasted for 1.5 sec, the [cAMP] and $k_{f,Raf}$ elevations for 1 min. To model activation of synaptic and nuclear kinases by these stimuli and consequent increases in GPROD and W, the equations of [33] were used with one modification. To sustain a significant value of W in the low stable state, prior to tetanic stimulation, a small basal rate constant $k_{W,bas}$ (0.001 min$^{-1}$) was added to the right-hand side of Eq. 18 of that paper.

To model the positive feedback from increases in W to the amplitude of ongoing synaptic activity and synaptic $Ca^{2+}$ transients, Eqs. 9-13 were implemented as described above ($N_{syn} = 1$). For numerical stability the time step was decreased to 1.5 msec. Every 100 time steps (150 ms), $Ca^{2+}_{syn}$ was elevated for 20 time steps. The frequency of this ongoing activity (6.67 Hz) is similar to theta rhythm. Ongoing synaptic activity also elevated $Ca^{2+}_{nuc}$ via Eq. 12, and elevated [cAMP] and $k_{f,Raf}$. [cAMP] and $k_{f,Raf}$ elevation amplitudes were increasing functions of W,

$$[cAMP] = \frac{W}{W+K_{max}}\left(0.15 \ \mu M\right) \tag{15}$$

$$k_{f,Raf} = \frac{W}{W+K_{max}}\left(0.24 \ min^{-1}\right) \tag{16}$$

## RESULTS

### Simulation of electrically induced late LTP

The reduced model qualitatively simulates changes in synaptic weight and the time course of the synaptic tag observed following tetanic stimulation of CA3-CA1 synapses (Fig. 2A-B). Three 1-sec tetani were simulated (stimulus parameters in Methods) with 5-min intervals between tetani. The TAG variable rises rapidly and decays with a time constant of a little over 3 hrs (Fig. 2A). Empirically, the tag lasts 2-3 hrs [31,32]. The level of gene product protein necessary for L-LTP, GPROD, declines over ~ 6 hrs. The synaptic weight W is increased at a rate proportional to the product, or overlap, of GPROD and TAG (Eq. 5) and Fig. 2B illustrates the time course of this overlap, denoted OV-P. L-LTP nears completion in 2-3 hr (time course of $W_{(tetanic)}$ in Fig. 2B). Similarly, experimental induction of L-LTP with BDNF (bypassing E-LTP) requires ~2 hr [56]. The increase in W is 140%, similar to the experimental EPSP increase observed after three or four 1 sec, 100 Hz tetani [57,58]. Parameters are at standard values (Table 1), except $k_{syn} = 25,000 \ hr^{-1}$.

Figure 2B also illustrates that the model yields substantial L-LTP for a theta-burst stimulus (TBS) (time course of $W_{(TBS)}$). TBS was simulated with a 2.5-sec $Ca^{2+}$ elevation (Methods).

In the model, the rate of TAG synthesis increases very steeply with $Ca_{syn}$ ($S_T^4$, with $S_T$ given in Eq. 2) and the rate of GPROD synthesis increases steeply with $Ca_{nuc}$ ($S_P^4$, with $S_P$ given in Eq. 4). If $Ca_{syn}$ and $Ca_{nuc}$ are fixed near $Ca_{bas}$ at 0.05 $\mu M$, $S_T$ is low (2.6 x $10^{-5}$) as is $S_P$ (4.8 x $10^{-5}$). For even lower $Ca^{2+}$, $S_T$ and $S_P$ are essentially zero. Thus, for $Ca^{2+}$ levels near or below $Ca_{bas}$, synthesis of synaptic tag and gene product is negligible. For constant $Ca_{syn}$ and $Ca_{nuc}$ levels, in order to achieve an equilibrium value of 0.1 or greater for TAG and GPROD, $Ca_{syn}$ and $Ca_{nuc}$ would have to equal or exceed 0.32 $\mu M$ and 0.23 $\mu M$, respectively. During the simulated tetani of Fig. 2, maximal values of TAG and GPROD are respectively near 1.0 and 0.2.

In the protocol of Frey and Morris which demonstrated synaptic tagging [32], two synapses converging onto overlapping populations of CA1 neurons were stimulated. L-LTP of synapse 1 was induced by three 1-sec tetani. Thirty-five min later, anisomycin was applied to block protein synthesis. Three tetani were then given to synapse 2. One hour separated the first tetani to synapses 1 and 2. Despite anisomycin, synapse 2 underwent L-LTP. If the prior tetani to synapse 1 were omitted, synapse 2 did not undergo L-LTP. These experiments suggested that the synapse 2 tetani set a synaptic tag. The tag enabled synapse 2 to "capture" proteins synthesized as a result of prior stimulation of synapse 1 and distributed within the dendrites of the common postsynaptic neurons.

The reduced model simulates this protocol (Fig. 3). Synapse 1 was given three tetani. To simulate anisomycin application, synthesis of the gene product protein GPROD was halted 35 min later. Further simulation details are in Methods. Synapse 2 was subsequently tetanized, elevating $TAG_2$. Although GPROD synthesis was blocked, much of the GPROD induced by the synapse 1 tetani was still present. This GPROD acts with $TAG_2$ to induce L-LTP of synapse 2.

In the reduced model, synaptic tagging and gene induction are completely dependent on $Ca^{2+}$ elevation (Eqs. 1-4). In contrast with the detailed model [33], the reduced model does not represent some processes ($Ca^{2+}$–independent kinase activation) that appear necessary for induction of L-LTP by chemical stimuli (forskolin or BDNF application). Imposed elevations of TAG and GPROD synthesis rates are the only way to simulate chem-LTP with the reduced model. The model of [33], which included $Ca^{2+}$-independent signaling, predicts that a small but significant rise in $Ca^{2+}$ is essential for the induction of chem-LTP. That model also simulated tetanic and theta-burst L-LTP and synaptic tagging (details in [33]). With that model, time courses of TAG, GPROD, and W are qualitatively similar to Figs. 2-3, except that the rise of TAG and GPROD to peak takes longer (~20 – 60 min) [33]. The lag is required for activation of kinases and phosphorylation of their substrates.

**Simulated memory maintenance requires ongoing synaptic activation**

Can the reduced and detailed models simulate persistent strengthening of synapses? To address this question, a representation of ongoing synaptic activity was added to the models. Activity was hypothesized to reactivate more strongly those synapses which have been strengthened, periodically elevating $Ca^{2+}$ to high levels at those synapses, reinforcing L-LTP. Ongoing activity was represented by brief $Ca^{2+}$ transients. Transients were assigned a random uniformly distributed amplitude (Eq. 13) and occurred at theta rhythm frequency. To selectively reinforce stronger synapses, the mean amplitude of the $Ca^{2+}$ transients was assumed to increase with synaptic weight. The argument supporting this assumption is given in the Discussion. For equations, see Methods (Eqs. 9-13). Initially the ongoing activity was assumed to be continuous. However, in subsequent simulations, episodic activity separated by inactive intervals was considered. Episodic activity may better represent synaptic reactivation during recall or sleep.

Reinforcement of a single synapse was first simulated. Bistability of synaptic weight was observed. A substantial separation between strong and weak weights occurred, and either state was preserved indefinitely during ongoing activity. A large imposed $Ca^{2+}$ transient could switch the weight from the low to the high state. Bistability relies on the positive feedback loop that operates as follows. $Ca^{2+}$ influx due to ongoing activity maintains a tonic, average level of activation of biochemical pathways responsible for L-LTP, represented in the model by a tonic elevation of GPROD and TAG. As a result, a tonic rate of L-LTP occurs, counteracting passive decay of synaptic weights to a basal value. L-LTP of synapses, in turn, leads to enhanced $Ca^{2+}$ influx at those synapses during ongoing synaptic activity. This reciprocal reinforcement of L-LTP and $Ca^{2+}$ influx constitutes a positive feedback loop. Figure 4 illustrates bistability with a bifurcation diagram for the case of a single synapse with $Ca_{nuc}$ and GPROD fixed and with ongoing activity. The bifurcation parameter $k_{ltp}$ is the rate constant that multiplies the overlap of TAG and GPROD to give the rate of L-LTP induction (Eq. 5). An increase in $k_{ltp}$ increases the strength of the positive feedback. As $k_{ltp}$ increases, a single stable state of low synaptic weight bifurcates to two stable states of high and low weight. In the high-W state, TAG and time-averaged synaptic $[Ca^{2+}]$ are substantially elevated (Fig. 4). The peak value of W attained after a tetanic stimulus without ongoing synaptic activity (Fig. 2 above) is well below the high-W state maintained by the activity-based positive feedback loop. Bistability similar to Fig. 4 could maintain a specific memory encoded by a network of strong synapses.

Figure 5A illustrates switching of the state of a synapse by an L-LTP event. At $t = 5$ hrs, three tetani are simulated. As W increases, the amplitude of synaptic $Ca^{2+}$ transients due to ongoing activity increases. The red trace is the time course of synaptic $Ca^{2+}$ with transients filtered by averaging over a moving time window of 36 sec (1,000 simulation time steps) and vertically scaled (multiplied by 4.0). As synaptic $Ca^{2+}$ increases, the post-tetanus decline of the synaptic tag variable is reversed. GPROD also increases to a plateau. Therefore, L-LTP is reinforced, and over ~15 hrs W climbs to the high stable steady state. If synaptic $Ca^{2+}$ transients are stopped, W passively decays to the low state (not shown). The first-order rate constant $k_{ltdbas}$ (Eq. 5) governs this decay. Simulations of bistability with synaptic activity (Figs. 4-5) use the same parameter values as simulation of L-LTP without ongoing activity (Fig. 2) with one exception. In Fig. 2, the rate constant for GPROD synthesis, $k_{syn}$, was higher (25,000 $hr^{-1}$). This higher value could represent a homeostatic upregulation of the ability to express L-LTP in the absence of ongoing activity.

It is plausible that *in vivo*, ongoing synaptic activity involved in memory consolidation and maintenance is episodic rather than continuous. Episodic activity could be associated with conscious recall, and/or with sleep. Empirically, the durations of such episodes, and the intervals between them, have not yet been well characterized, especially over substantial time periods in behaving animals. Therefore, to simulate episodic activity, periods of $Ca_{syn}$ and $Ca_{nuc}$ elevations at theta frequency were simply separated by inactive intervals with basal $Ca_{syn}$ and $Ca_{nuc}$ (0.04 μM). The reduced model simulates induction and maintenance of high synaptic weight with episodic activity (Fig. 5B). Here, 15-sec active periods are separated by 45-sec inactive intervals. Three tetani again induce a transition to stable, high W. The strength of positive feedback had to be increased to allow this transition ($k_{ltp}$ was increased to 1.0 $hr^{-1}$). With episodic activity, GPROD and average $Ca_{syn}$ are much lower in the high-W state. The activity/inactivity cycle time could be increased from 1 min to 5 or 10 min while preserving the transition from low to high W, as long as the active cycle fraction remained at ~ 0.25. Longer cycle times (several hrs) disrupted the state transition. However, once W settled in the high state, it was more resistant to inactivity. For example, with model parameters as in Fig. 5B, W could be maintained in a high state (with fluctuations) given a cycle time of 24 hrs and an active cycle fraction as low as 0.06.

For the bistability illustrated in Figs. 4-5 to be a plausible candidate for a memory storage mechanism, it must be robust to changes in parameter values. Robustness was examined by a simulations in which the model parameters (Table 1) were successively varied by +30% and -30%. The effect of these variations on both the high and low steady states of W was examined. There are 15 parameters (leaving out $W_{inc}$ and $W_{max}$, which are only used in Fig. 8). Therefore, 62 simulations were carried out (four for each parameter, and two controls with all parameters at standard values in Table 1). The resulting scatter plot (Fig. 6A) shows that most parameter variations do not significantly change the high and low values of W. Most of the 31 points are clustered very near the red control point corresponding to standard parameter values (Table 1).

With one exception, none of the changes of W from the control values are greater than 40%, suggesting the model is reasonably robust. The two points with substantial decreases of W-high from control correspond to a 30% increase in $K_{SP}$ (Eq. 4) and a 30% decrease in $f_{nuc}$ (Eq. 12). The decrease in $f_{nuc}$ eliminates the upper state, so that W falls to W-low. These points illustrate that W-high is particularly sensitive to a decrease in $[Ca_{nuc}]$ (due to a decrease in $f_{nuc}$) and that W-high is sensitive to a decrease in GPROD induction at fixed $[Ca_{nuc}]$ (due to an increase in $K_{SP}$). An increase in $K_{SP}$ decreases the effect of a given $Ca_{nuc}$ level on GPROD synthesis, by decreasing $S_P$ in Eq. 4. The points with significant changes in W-low from control correspond to 30% changes in $k_{ltpbas}$ and $k_{ltdbas}$ (Eq. 5). These small basal rates of synaptic potentiation and depression determine W-low, but not W-high.

As discussed in Methods, the rates of synthesis of TAG and GPROD are extremely steep, supralinear functions of $[Ca_{syn}]$ and $[Ca_{nuc}]$ respectively. Is the bistability illustrated in Fig. 4 is robust to decreases in the supralinearity of these relationships? Equations 1 and 3 were modified by removing one sigmoidal function of $[Ca^{2+}]$ from the right-hand sides. A power of $S_T$ was removed from Eq. 1, and a power of $S_P$ was removed from Eq. 3. With these changes, bistability was preserved. Figure 6B illustrates the corresponding bifurcation diagram, which is very similar to Fig. 4. This diagram was also robust to parameter variations like those in Fig. 6A (not shown).

I also examined the effect of replacing the steep dependence of TAG and GPROD synthesis rates on $[Ca^{2+}]$ with step functions. The rate of TAG synthesis was assumed to switch from 0 to $k_{phos}$ at $Ca_{syn} = 0.7$ μM, and the rate of GPROD synthesis switched from 0 to $k_{syn}$ at $Ca_{nuc} = 0.15$ μM. A bistable diagram similar to Fig. 4 was again obtained. Below a critical $k_{ltp}$ value (~0.2), only the lower state was stable, with the same W value as in Fig. 4. Thus, the steep $Ca^{2+}$ functions in Eqs. 1 and 3 ($S_T^4$, $S_P^4$) are functionally similar to abrupt switches.

If peak values of $Ca_{syn}$ or $Ca_{nuc}$ are decreased during tetanic simulation or during ongoing activity, then peak and steady-state values of W decrease. However, the lack of any negative feedback loop in the model implies oscillatory dynamics will not be observed, irrespective of values assumed for $Ca_{syn}$ or $Ca_{nuc}$. During optimization simulations in which stimulus and model parameters were varied, oscillations in W or other variables were never seen. In addition, for no parameter values did I observe more than two stable steady states of W. The single positive feedback loop of the mode does not seem able to sustain more than two stable states.

For Figs. 3-6, ongoing synaptic activity was represented by brief $Ca^{2+}$ elevations occurring at a frequency characteristic of theta rhythm (5.6 Hz). A recent study [59] found that activities of single neurons in various regions of the human brain are often phase locked to theta oscillations. Rodent hippocampal place cells, when active, have long been observed to fire in phase with theta [60]. However, the recent human data [59] also showed that neuron activity is often phase locked to gamma frequency (30-90 Hz), and that hippocampal neurons sometimes fire at delta frequencies. In rodents, so-called sharp waves are associated with collective firing of hippocampal neurons [61]. Sequences of hippocampal place cells activated during experience are sometimes replayed during sharp waves [62], suggesting that bursts of activity during sharp waves may play a role in memory consolidation. The representation of ongoing activity used for Figs. 4-6 is not inclusive of the variety of neuronal activation dynamics that may occur during ongoing synaptic activity necessary for memory consolidation and maintenance.

To begin to examine whether the model dynamics are preserved with other representations of ongoing activity, the simulations of Figs. 4 and 5A were repeated for other frequencies of $Ca^{2+}$ elevation: 28 Hz, 14 Hz, and 2.8 Hz. The latter frequency is similar to delta rhythm. With the simulation time step of 36 msec, these frequencies correspond respectively to $Ca_{syn}$ and $Ca_{nuc}$ elevations at every time step (continuous elevation), every second time step, and every 10th time step. Other stimulus and model parameters were as for Fig. 5A. A bifurcation diagram similar to Fig. 4 resulted for all frequencies. The value of $k_{ltp}$ at which the upper state disappeared decreased as frequency increased. For 28, 14, 5.6, and 2.8 Hz; this value was respectively 0.008, 0.01, 0.014, and 0.016 hr⁻¹. Furthermore, for all activity frequencies, time courses of transitions from low to high synaptic weight were similar to Fig. 5A, in shape and duration.

## Bistability and maintenance of synaptic strength are also simulated by the detailed model of L-LTP

To test whether the dynamics illustrated in Figs. 2-6 are peculiar to the reduced model, ongoing synaptic activity was added to the previous detailed model of L-LTP induction [33]. That model represents stimulus-induced elevation of $Ca_{syn}$, $Ca_{nuc}$, and cAMP, activation of kinases, phosphorylation of a synaptic tag and transcription factors, gene product induction, and increase of a synaptic weight W. The differential equations of that model were supplemented with Eqs. 9-13, which implement ongoing synaptic activity (brief $Ca_{syn}$ elevations) and positive feedback between increased W and increased $Ca_{syn}$ elevations. Further details are in Methods.

This extended model exhibits bistability in W. Induction of L-LTP causes a permanent state transition from low to high W. Figure 7 illustrates the dynamics of this model. Initially W is low, and the strength of ongoing activity is low (small amplitude of $Ca_{syn}$ transients in Fig. 7A for $t < 60$ hrs). At $t = 60$ hrs, three tetani are given at 5-min intervals (stimulus parameters in Methods). The large elevation of $Ca_{syn}$ (Fig. 7A) and $Ca_{nuc}$ activates synaptic and nuclear kinases (CaM kinases II and IV, MAP kinase) (Fig. 7B). Activation of kinases leads to induction of TAG and GPROD (Fig. 7B, 7C). As in the reduced model, the product or overlap of TAG and GPROD drives an increase in W (Fig. 7C). Positive feedback from W to $Ca_{syn}$ increases the amplitude of the ongoing $Ca_{syn}$ transients (Fig. 7A for $t > 60$ hrs). Consequently kinase activation is reinforced, TAG and GPROD increase to high plateaus, and W is maintained in the upper state. The upper state of W in Fig. 7C is above 1 because in the detailed model, the increase in W is not limited by an imposed upper bound of 1 (as in Eq. 5) but rather by depletion of a synaptic protein (see [33] for details).

The robustness of the steady states of W to parameter variation was assessed. Each parameter in the detailed model was varied by +30% and -30%. The changes in the steady-state values of W were always less than 55%. Therefore, bistability in the detailed model appears reasonably robust to parameter variation.

## Simulated loss of memory by reduction of ongoing activity

The simulations of bistability in Figs. 4-7 do not represent one of the most important elements of biological memory maintenance – the ability to weaken strong synapses and forget memories. In a network of synapses like the single synapse of Fig. 5, synapses would never be removed from the strong state. Ongoing L-LTP events would drive an increasing proportion of synapses into that state, eventually erasing all specific patterns of strengthened synapses, and eliminating all memories.

To simulate aspects of memory loss as well as maintenance, bistability of synaptic weights due to ongoing activity was implemented in a small group of 5 synapses, with the reduced model (Figs. 1A-B). Synaptic weakening and memory loss was also implemented. Periodic reductions of ongoing activity, in concert with repeated brief activity spikes (single tetani), were hypothesized to respectively induce loss and reformation of synaptic strength. Convergence onto a single postsynaptic neuron was assumed, similarly to other studies (*e.g.*, [63]). In the reduced model, the postsynaptic neuron is only represented by two variables. These are the nuclear $Ca^{2+}$ concentration $Ca_{nuc}$, and the concentration of induced gene product GPROD. Therefore, a common $Ca_{nuc}$ and a common GPROD were coupled identically to all the synapses. For further details see Methods (Fig. 1B, Eqs. 6-14). Memory formation and maintenance corresponds to persistent strengthening of a subset of synapses in the group. Each strengthened synapse was conceptualized as storing, or participating in the storage of, a distinct memory. Thus, weakening of one of the five synapses corresponds to loss of a memory.

The group dynamics were simulated for 20,000 hrs (833 days). The five synaptic weights were initialized to 0.1, near the low steady state of W. Single tetani were given every 500 hrs, cycling repeatedly through synapses 1-5. The first tetanus placed synapse 5 in the high weight state, near 1. Subsequently, synapses 1 through 4 underwent tetanus and entered the high weight state. Then, at synapses 1-4, the frequency of synaptic reactivation was periodically reduced.

Initially, synaptic transitions from high to low weight were only induced if ongoing activity was virtually stopped. A >99% reduction in the frequency of $Ca^{2+}$ elevations was required. This behavior is explained as follows. First, for single tetani to be able to induce low to high transitions, the rate constant governing positive feedback ($k_{ltp}$) had to be set to a relatively high value (0.4 hr$^{-1}$, well to the right of the range in Fig. 4). Then, with high $k_{ltp}$, the high-W state became very resistant to perturbations, and virtually complete cessation of reactivation was required to return W to the low state.

To allow for transitions from high to low weight with a somewhat smaller reduction in reactivation frequency, it was necessary to introduce a simple representation of synaptic scaling, so that the value of the high weight state decreased as the number of strong synapses increased. The scaling is described in Methods (Eq. 14). As the high weight state is scaled down, the amplitude of ongoing synaptic $Ca^{2+}$ transients decreases (Eqs. 9-10). Therefore, as the number of strong synapses increases, positive feedback between synaptic $Ca^{2+}$ and W is weakened. The high weight state becomes more susceptible to destabilization by decreased synaptic activity.

Then, the frequency of synaptic reactivation was periodically reduced at synapses 1-4 as follows. At any given time, for exactly one synapse, the fraction of theta rhythm cycles inducing a transient elevation of synaptic $Ca^{2+}$ was reduced from 1.0 to 0.1. Even with scaling implemented, a 90% reduction in reactivation was required to consistently destabilize the high weight state. Every 1,200 hrs, the identity of the deactivated synapse changed, cycling repeatedly through synapses 1-4. Synapse 5, in contrast, was always activated every theta cycle. The result was simulated selective forgetting as well as selective permanent memory storage during a long, 20,000-hr simulation (Fig. 8). Memories stored by synapses other than synapse 5 were repeatedly "forgotten", *i.e.*, these synaptic weights switched repeatedly from the high to the low state. The decay of synaptic weight occurred on a time scale of a few days, governed by the first-order rate constant in Eq. 14, $k_{ltdbas}$ = 0.01 hr$^{-1}$. Synapses 1-4 were thereby repeatedly made available for storage of new memories, and were then repeatedly strengthened by L-LTP – inducing events. Synapse 5, with no imposed decreases in the ongoing activity, never showed forgetting. Figure 8A illustrates repeated forgetting at a representative synapse, synapse 4, and permanence of memory at synapse 5. The selective preservation of synapse 5's strength suggests a possible mechanism for preservation of an important memory for as long as a lifetime.

Figure 8B illustrates in more detail memory formation and forgetting at another synapse, synapse 2. $Ca_{syn,2}$ levels and $TAG_2$ are higher when $W_2$ is in the upper state. However, fluctuations in $Ca_{syn,2}$ levels and $TAG_2$ also occur between $W_2$ state transitions. These fluctuations are driven by transitions of other synapses between states of high and low W.

## DISCUSSION

### A reduced model of L-LTP induction can represent the role of essential biochemical nonlinearities

The reduced model (Fig. 1A) represents synthesis of a synaptic tag and a gene product (variables TAG and GPROD) as very steep, "threshold" functions of synaptic and nuclear $Ca^{2+}$ respectively. In turn, TAG and GPROD multiply together to give the rate of increase of a synaptic weight W. The reduced model, like the detailed model it is derived from [33], was developed to represent L-LTP in the hippocampal CA3-CA1 pathway. Cortical regions, rather than the hippocampus, are likely to predominate in long-term storage of memories (months or longer). However, it is plausible that the essential elements of these models – synaptic tagging, gene induction, and reinforcement of strengthened synapses by ongoing neuronal activity – qualitatively describe L-LTP induction and memory maintenance at cortical as well as hippocampal synapses.

The reduced model qualitatively simulates the amplitude and time course of L-LTP and the time course of the synaptic tag following tetanic or theta-burst stimulation (Fig. 2). As detailed in Methods, the steepness of the relationships between $Ca^{2+}$ levels and synthesis rates of TAG and GPROD constitutes a concise representation of the convergence of multiple kinases to activate the synaptic tag and initiate gene expression. Activation of some individual kinases also depends supralinearly on $Ca^{2+}$, requiring binding of four $Ca^{2+}$ ions to calmodulin.

I argue such a steep dependence of L-LTP induction on synaptic and nuclear $Ca^{2+}$ will prove to be an essential feature of any model of L-LTP induction. Only a very steep dependence can explain the empirical observation that late LTP can be induced by three brief tetani, each generating a ~ 1 sec, 10-20 fold elevation of $Ca^{2+}$. Without supralinearity, a 20-fold elevation of $[Ca^{2+}]$ lasting for 3 sec would drive only a negligible increase in a variable such as synaptic strength. The much longer time constant of the latter variable would almost completely damp its response to the brief stimulus. In contrast, supralinearity acts to amplify the biochemical response to $Ca^{2+}$ elevation. As a concrete example, suppose the steep relationships between $Ca^{2+}$ elevation and synthesis rates of TAG and GPROD amplify the effect of a $Ca^{2+}$ transient by $10^4$. If a 20-fold, 2-sec elevation of $Ca^{2+}$ is multiplied by $10^4$, this corresponds to delivering a 2-sec stimulus of amplitude $\sim 10^5$ to a slow variable such as synaptic weight. Even if the latter variable has a very slow time constant of $\sim 10^5$ seconds (a day or longer), it would substantially increase.

The reduced model represents induction of gene expression (Eq. 3) and capture of the resulting gene product by tagged synapses (Eq. 5). Empirically, local dendritic translation of proteins may provide an essential intermediate step in long-term synaptic strengthening. The time required for transcription, translation, and movement of proteins or mRNAs from the nucleus to the synapse is likely to be a few hours or more in many cases. For example, NMDAR subunits are transported at ~120 μm hr$^{-1}$ in dendrites, so that in 1 hr, a subunit could move about the length of a dendritic branch [64]. The synaptic tag may capture locally translated proteins, sustaining the first few hrs of L-LTP [65]. However, over longer times, the strengthened synapses must also preferentially capture proteins resulting for transcription, which is required for the late phase of L-LTP [29]. Thus one can envision a second, long-lasting synaptic "tag", which functions to capture the products of transcription.

## The reduced and detailed models simulate memory maintenance by ongoing synaptic reactivation

It is plausible that biochemical pathways reactivated by ongoing activity in synaptic networks mediate repeated L-LTP events, preserving strengthened synapses. However, a conceptual picture has not yet been developed describing how L-LTP might interact with ongoing neuronal activity to create multiple stable states of synaptic weights. With the reduced model of L-LTP induction, synaptic weight dynamics were simulated in the presence of ongoing neuronal activity, represented by brief Ca$^{2+}$ elevations occurring with a frequency similar to theta rhythm. These elevations were assumed to increase with synaptic strength. Simulations illustrated bistability of synaptic weight. Induction of L-LTP could switch a synapse from a "low" to a "high" weight, and ongoing activity maintained the synapse in the high state. These simulations (Figs. 4-5) demonstrate synaptic reentry reinforcement (SRR) [23,24]. Bistability relies on positive feedback between synaptic strengthening and increased Ca$^{2+}$ elevations. The bifurcation diagram of Fig. 4 illustrates the emergence of bistability as the strength of positive feedback is increased. Figure 5 illustrates that following a single L-LTP event, SRR is induced, so that the synapse slowly strengthens over ~1 day to a plateau. Assuming that persistent strengthening of a synapse corresponds to storage of a very simple memory, this bistability allows for memory storage. Long-term memory storage based on bistability of connection strength has been empirically implemented in a VLSI circuit [66].

The simulated high and low weight states were found to be robust to moderate (30%) changes in parameter values (Fig. 6A) with the exception that a $f_{nuc}$ decrease eliminated the high state. Bistability was also robust to a decrease of supralinearity in the equations describing induction of synaptic tag and gene product by Ca$^{2+}$ (Fig. 6B). The empirical relationships between Ca$^{2+}$ elevation and induction of TAG and GPROD might not be as sharp as in Eqs. 1-4. If so, Fig. 6B suggests positive feedback between synaptic weight and [Ca$_{syn}$] elevations is still a plausible mechanism for sustaining bistability.

The hypothesis that reverberatory activity in synaptic networks maintains stored memories would appear to imply that ongoing activity generates higher Ca$^{2+}$ elevations at strong synapses, thereby selectively reactivating biochemical pathways for strengthening those synapses. The models presented here <u>require</u> this assumption (details in Methods). Its justification depends on the following argument. During ongoing activity, a casual "pre before post" neuronal firing order is likely to occur more frequently at strong synapses, because these synapses play a larger role in driving postsynaptic neuronal firing. In turn, a "pre before post" Hebbian firing order would more strongly elevate synaptic Ca$^{2+}$, because pairing of presynaptic glutamate release with subsequent postsynaptic depolarization maximizes Ca$^{2+}$ influx through NMDA receptors. Alternatively, dendritic spines corresponding to stronger synapses are likely to have a higher surface density of NMDA receptors and voltage-gated $Ca^{2+}$ channels. These characteristics would generate larger elevations of $[Ca^{2+}]$.

Bifurcation diagrams similar to Fig. 4, and state transitions similar to Fig. 5A, were also seen when the frequency of ongoing activity ($Ca_{syn}$ and $Ca_{nuc}$ elevations) was varied from 28 to 2.8 Hz. This preservation of bistability for different frequencies was expected, because the positive feedback underlying bistability is based on the qualitative hypothesis that a synaptic weight increase corresponds to an increased amplitude of ongoing synaptic $Ca^{2+}$ elevations (Methods, Eqs. 9-10). The specific frequency of $Ca^{2+}$ elevations, and their detailed shape, is of less importance. For example, an increased amplitude of $Ca^{2+}$ bursts driven by repeated sharp waves could also contribute to such positive feedback.

*In vivo*, synaptic activity is more likely to be episodic than continuous. Specific patterns of synapses, which may correspond to memory traces, are hypothesized to be activated in episodes of conscious recall, during sleep, or during spontaneous awake activity. For the reduced model to be compatible with the SRR hypothesis, episodic ongoing activity should therefore suffice to maintain strong synapses. These dynamics were observed (Fig. 5B). Bistability, and permanent synaptic strengthening by tetani, were seen when the ongoing activity consisted of brief periods of $Ca^{2+}$ elevations, at theta-rhythm frequency, with the periods separated by intervals (timescale of minutes) without activity. Furthermore, once a synapse was in the stable high weight state, episodic activity could maintain that state even if the inactive intervals were relatively long (~1 day). Inactive intervals cannot be too long, however, The first-order rate constant $k_{ltdbas}$ (Eq. 5) represents the time scale of passive synaptic weight decay (100 hr for $k_{ltdbas}$ as in Table I). Intervals of inactivity longer than this time scale would return synapses to the low weight state, erasing any associated memories.

Bistability, and tetanus-induced state transitions, were also seen when the more detailed model of L-LTP induction was combined with ongoing synaptic activity (Fig. 7). Similar dynamics are likely to occur in any model with the following elements: 1) strongly supralinear coupling between $[Ca^{2+}]$ and synaptic weight increase, 2) ongoing background synaptic activity represented by $[Ca^{2+}]$ elevations, and 3) positive feedback in which synaptic weight increase elevates the amplitude of $[Ca^{2+}]$ transients.

## Biochemical bistability alone may not be able to sustain long-term memory storage

Previous models (see Introduction) have postulated that bistability in synaptic strength could be generated by biochemical positive feedback loops involving kinase activation or mRNA translation. Those positive feedback loops rely implicitly on ongoing neuronal activity to sustain $Ca^{2+}$ influx and consequent activation of the MAPK cascade [41] and/or PKA. However, those feedback loops do not require SRR, because SRR postulates greater reactivation and $Ca^{2+}$ influx at strong synapses, whereas biochemical feedback might generate bistability given a constant ongoing $Ca^{2+}$ influx.

A recent investigation suggests biochemical bistability alone cannot underlie biological memory maintenance. Fusi et al. [16] argued a model based on bistability of synaptic weight states cannot provide a biophysically realistic model of memory storage, because in a neuronal network, ongoing neuronal activity combined with molecular turnover would randomly depotentiate individual strong synapses, and rapidly degrade a memory even if it was comprised of a large number of synapses ($>10^6$). This argument may not, however, suffice to rule out a bistable SRR-based memory maintenance mechanism. A synapse, defined as the active contact area between a pair of neurons, is commonly comprised of a considerable number of individual synaptic contacts between specific axonal boutons and dendritic spines. For example, on the order of 10 individual contacts have been estimated to characterize excitatory synapses between CA3 pyramidal cells [67], CA3 and CA1 pyramidal cells [68] or neocortical pyramidal cells [69]. Each individual spine and synaptic contact is a separate biochemical unit. If one of these units spontaneously switches to a weak state, this might not significantly alter the ongoing neuronal activity pattern, as long as each synapse is comprised of multiple individual contacts. SRR may therefore continue to operate, reactivating synapses where the majority of contacts are strong. The reactivation may re-strengthen those individual contacts that have weakened. In this way, the memory may be preserved. Therefore, for networks in which synapses are generally comprised of multiple individual contacts, the argument of Fusi et al. [16] may not suffice to rule out bistability mediated by SRR as a viable biological memory storage mechanism. Further analysis of this question is important.

**Simulated selective forgetting relies on less frequent reactivation of memory**

The dynamics illustrated in Figs. 5 and 7 do not yet represent a satisfactory system for biological memory storage, because those dynamics have no mechanism for reversing L-LTP and forgetting memories. Some forgetting may rely on switching strong synapses to a weak state by long-term depression (LTD). However, it is not evident how LTD could favor selective forgetting of memories that are old and/or no longer relevant for the life of an animal. I therefore suggest selective preservation of memories may be based on more frequent reactivation of groups of synapses that store memories relevant to current experience. For example, current sensory experiences may generate similar inputs to cortex or hippocampus as did similar remembered experiences. These inputs would reactivate some of the same synapses that were activated when those memories were formed. Conversely, memories dissimilar from current experience might be less frequently re-activated and susceptible to loss.

To simulate selective maintenance of memories, the reactivation frequency was periodically varied for each synapse in a group of 5 synapses. Each strengthened synapse was envisioned as participating in the encoding of a distinct memory. The resulting simulation (Fig. 8) suggests the model may represent a mechanism for memory storage and forgetting. During the very long simulated time (> 2 yrs), "memories" were formed due to L-LTP, stabilized by ongoing synaptic activity, and forgotten when the reactivation frequency diminished. At one synapse, the memory was permanently preserved when the reactivation frequency was kept high.

L-LTP lasting for months has been observed in rat dentate gyrus [70]. Intriguingly, this L-LTP could be reversed by repeated exposure to an enriched environment beginning 14 d post-tetanus. This result may support the hypothesis that forgetting involves less frequent reactivation of synapses storing memories dissimilar to current experiences. During environmental enrichment, it is likely that the pattern of ongoing synaptic activity would shift, such that some previously strengthened synapses would be less active and undergo less reinforcement of their synaptic weights. Therefore, those synapses might not be maintained in a strong state, and corresponding memories would be lost.

## Synaptic weight maintenance by episodic activity, and experimental predictions

If a positive feedback loop between synaptic strengthening and activity-driven $Ca^{2+}$ elevation does selectively activate biochemical pathways at strong synapses, reinforcing L-LTP at those synapses, then experimental predictions follow. The synaptic tag identified as being necessary for L-LTP [31,32] should be persistently and selectively set at strong synapses due to ongoing neuronal activity, *in vivo* or in active slice preparations. Once the molecular correlate of the tag is identified, this prediction can be tested. The expression of some genes necessary for L-LTP should be downregulated when neuronal activity is inhibited. The hypothesis that activity-induced elevations of synaptic $Ca^{2+}$ are higher at strengthened synapses might be tested in cultures of active, interconnected cortical or hippocampal neurons. LTP of a specific synapse might be induced by simultaneously injecting current into the pre- and postsynaptic neurons. Concurrently, $[Ca^{2+}]$ in spines of the postsynaptic neuron could be estimated by use of fluorescent indicators. If spines corresponding to the synapse in question can be identified, the hypothesis predicts time-averaged $[Ca^{2+}]$ in those spines would be seen to increase after LTP.

If ongoing neuronal activity preferentially activates strong synapses by means of a "pre before post" neuronal firing order, then experiments may reveal repeated occurrences of specific spatiotemporal sequences of neural firing. For example, if neurons A, B, and C are successively linked by strong synapses, then the firing order A->B->C may repeatedly occur *in vivo*. The concept of a repeated, ordered sequence of neuronal firing is commonly termed a synfire chain. Indeed, experiments have recently demonstrated synfire chains *in vivo* in cat visual cortex [71]. Earlier experiments suggested synfire chains operate in primate frontal cortex *in vivo* [72] and a number of modeling studies have investigated the dynamics of synfire chains [73]. Thus, ongoing reverberatory activity does appear to characterize at least some *in vivo* synaptic networks. Although maintenance of specific memories may correlate with repeated sequential activation of synaptic patterns or synfire chains, an important caveat is in order. Commonly, large response variability to repetitions of the same stimulus is seen at the single neuron level, for example in hippocampal place cells [74], and in visual cortex [75] (for an exception see [76]). Such dynamic variability suggests that reactivation of a synaptic memory pattern or trace will not always result in the same order of firing of individual neurons. Often only part of the trace may be reactivated. However, the SRR hypothesis predicts that for any specific strong synapse in a maintained trace, reactivation <u>must</u> sometimes induce LTP of that synapse. Thus some, but not necessarily all, reactivations are expected to yield the "pre before post" firing order that correlates with LTP.

The models presented here do not include inhibitory synapses. With periods of decreased activity to return synapses to the low weight state, inhibition is not necessary to prevent all excitatory synapses from eventually converging to a state of high synaptic weight. However, as memory maintenance in specific neuronal systems is modeled, it will be important to incorporate both ongoing and variable synaptic activity, and investigate the effects of both recurrent circuits and inhibitory synapses. It will also be necessary to examine the memory capacity of larger networks of synapses incorporating SRR. This might be pursued by implementing SRR in some current network models, including models for hippocampal or cortical learning, as well as less specific models that represent types of learning, such as sequential memory [77].

Late long-term depression (L-LTD) can be induced at hippocampal synapses by low-frequency stimulation, and is protein synthesis-dependent [78,79]. Like L-LTP, L-LTD relies on synaptic tagging and capture of plasticity factors [80]. Ongoing synaptic activity could induce L-LTD at strengthened synapses, which would compete with L-LTP. In that case, memory maintenance based on ongoing activity could only occur if L-LTD was less strong, on average, than ongoing reinforcement of L-LTP. I am currently attempting to characterize constraints on L-LTD kinetics that allow simulation of empirical data, but nevertheless do not negate positive feedback and bistability of synaptic weights. Details will be reported in future work.

Independently of the present model, if synaptic weight dynamics are altered by variations in ongoing neuronal activity that are largely due to changes in an animal's environment and experiences, then *in vivo* distributions of synaptic weights cannot be expected to converge to equilibrium. Therefore, investigations focused on equilibrium distributions (*e.g.*, [81]), although important, may not suffice to accurately predict weight distributions *in vivo* and determine whether proposed learning rules can represent *in vivo* weight regulation.

Further experimental and theoretical investigations that examine more quantitatively the dynamics of memory maintenance by synaptic reactivation will help to solve one of the major outstanding questions in neurobiology, the preservation of memories for up to a lifetime in animals including humans.

**Acknowledgements:** I thank Harel Shouval, John Byrne, and Anthony Wright for critically reading the manuscript.

## FIGURE LEGENDS

**Figure 1**. Schematic of the model for L-LTP induction. Electrical stimuli (tetanus or theta-burst) elevate synaptic and nuclear $Ca^{2+}$. Steep sigmoidal functions of synaptic $Ca^{2+}$ and nuclear $Ca^{2+}$ transduce $Ca^{2+}$ elevations into activation of synaptic tagging (elevation of the variable TAG) and activation of gene expression (elevation of GPROD). These sigmoidal functions are denoted $Th_1$ and $Th_2$. Equations are given in Methods. High synaptic $Ca^{2+}$ elevations are required to set the tag for L-LTP. L-LTP is simulated by increase of a synaptic weight W at a rate proportional to the product of TAG and GPROD. (B) Model of a group of synapses converging on a single postsynaptic neuron. At each synapse, ongoing activity is modeled with brief $Ca^{2+}$ transients at theta frequency (see Methods for details). Strengthening of any of the weights $W_1 \ldots W_N$ increases the amplitude of $Ca^{2+}$ transients at that synapse and the level of nuclear $Ca^{2+}$. In turn, increased synaptic $Ca^{2+}$ synergizes with increased nuclear $Ca^{2+}$ to strengthen the corresponding synaptic weight.

**Figure 2**. Simulation of tetanic and theta-burst L-LTP. (A) At $t = 2$ hrs, three tetani are simulated at 5-min intervals. Each tetanus is simulated by concurrent 1-sec elevations of synaptic and nuclear $Ca^{2+}$ to 0.9 μM and 0.45 μM respectively. $Ca^{2+}$ elevations induce synthesis of GPROD and TAG. The time course of GPROD is vertically scaled (multiplied) by 3.0. (B) The rate of synaptic weight increase is proportional to the product, or overlap, of TAG and GPROD (time course of OV-P). The change in W(tetanic) is L-LTP of ~140%. L-LTP induced by a theta-burst stimulus is of similar magnitude, as illustrated by the W(TBS) time course.

**Figure 3**. Simulation of the synaptic tagging protocol of [32]. Parameter values, including stimulus parameters, are as in Fig. 2. At $t = 2$ hr, synapse 1 is given three tetani at 10-min intervals, increasing $TAG_1$, GPROD, and $W_1$. Thirty-five min after tetani, GPROD synthesis is blocked, but GPROD remains elevated, decaying slowly. At $t = 3$ hr, synapse 2 is given three tetani. $TAG_2$ and $W_2$ are increased as a result. Time courses of GPROD, $W_1$, and $W_2$ are vertically scaled by 3.0.

**Figure 4**. A bifurcation diagram illustrates bistability of synaptic weight. The bifurcation parameter $k_{ltp}$ is the rate constant for increases in W during L-LTP. As $k_{ltp}$ increases past 0.014, a steady state of high W appears. In this state, the time average of synaptic $[Ca^{2+}]$ is elevated (running average taken over 10 min of $Ca^{2+}$ transients). Consequently the tag for L-LTP is set (TAG is near 1.0), allowing for continual reinforcement of L-LTP. Synaptic $[Ca^{2+}]$ is vertically scaled by 4.0. Parameters other than $k_{ltp}$ are at their standard values (Table 1).

**Figure 5**. Time courses of W, GPROD, TAG, and time-averaged synaptic $Ca^{2+}$ following 3 tetani to the single synapse of Fig. 4. Tetanus parameters are as in Fig. 2. Initially, W is in the low steady state of Fig. 4. The tetani elevate TAG, GPROD, and W. (A) With continuous background synaptic activity, TAG, GPROD, and W converge to the high steady state of Fig. 4. $k_{ltp} = 0.4$ hr$^{-1}$, well to the right of the range in Fig. 4, so the high state of W is close to 1. Synaptic $[Ca^{2+}]$ is vertically scaled by 4.0 and GPROD is scaled by 0.5. Other parameters are as in Figure 4. (B) Tetani also induce a transition to the high steady state with episodic, rather than continuous, background synaptic activity. Periods (15 sec) of brief $Ca_{syn}$ and $Ca_{nuc}$ elevations at theta-rhythm frequency are separated by intervals (45 sec) with no activity. All other parameters are as in (A) except $k_{ltp} = 1.0 \ hr^{-1}$.

**Figure 6**. Bistability is robust to parameter and equation changes. (A) Scatter plot with 31 points. The x-axis plots values of W in the low steady state of Fig. 5A, and the y-axis gives values of W in the high steady state. The control point (red square) corresponds to all parameters at standard values (Table 1). To generate the other points, the parameters were successively changed by plus or minus 30%. Most points are clustered very close to the control point. (B) Bifurcation diagram analogous to Fig. 4. To generate this diagram, one power of $S_T$ was removed from Eq. 1, and one power of $S_P$ was removed from Eq. 3. Parameter values are as in Fig. 4 except $k_{phos} = 3,330 \ hr^{-1}$, $k_{syn} = 220 \ hr^{-1}$.

**Figure 7**. Bistability of synaptic weight in the model of [33] with ongoing synaptic activity. (A) Time courses of synaptic and nuclear $Ca^{2+}$. $Ca_{syn}$ varies continuously (black envelope) due to the brief imposed elevations that represent synaptic activity. At $t = 60$ hrs, L-LTP is induced by tetanic stimulation (parameters in Methods). $Ca_{syn}$ is briefly elevated due to the tetanus (black vertical line). As W increases after tetanus, positive feedback between W and synaptic activity increases the amplitude of the $Ca_{syn}$ transients. $Ca_{nuc}$ increases less. (B) Time courses of synaptic variables –activities of CAMKII and MAPK, and TAG. The tetanus briefly elevates TAG, which at first starts to decline. However, the increase in W and the corresponding increase in $Ca_{syn}$ activate CAMKII and MAPK, which phosphorylate substrates and subsequently increase TAG. (C) Time courses of nuclear CAMKIV activity, GPROD, and W. In (B) and (C), time courses of CAMKII and GPROD are vertically scaled by 0.2 and 15.0, respectively. Parameter values are as in Table 1 and in [33] (their Table I) except for the following changes. From Table 1: $Ca_{tet} = 1.0$ $\mu M$, $Ca_{limit} = 1.0 \ \mu M$, $K_{syn} = 10.0$, $K_{sum} = 0.1 \ \mu M$. From [33]: $k_W = 0.22 \ \mu M^{-1} \ min^{-1}$, $\tau_W = 700$ min, $V_P = 0.02 \ \mu M^{-1} \ min^{-1}$, $\tau_P = 50$ min, $k_{f,Raf}$ (basal) = 0.

**Figure 8**. Simulation of maintenance and forgetting of simple memories, with each memory represented by strengthening of one of five synapses. L-LTP – inducing tetani (simulated as in Fig. 5) are periodically repeated at all synapses. The frequency of ongoing activity is periodically decreased at synapses 1-4, but is kept high at synapse 5. Synaptic scaling is implemented as described in Methods (Eq. 14). Synapses 1-4 repeatedly gain memory due to L-LTP, and also repeatedly lose their memory due to decreases in ongoing activity. In contrast, the memory at synapse 5 is preserved. (A) Time courses of synaptic weight for synapse 4 (blue trace) and for synapse 5. Model parameters are as in Table 1. Arrows, imposed decreases of activity at synapse 4, which decrease $W_4$. (B) Details of the dynamics at another synapse, synapse 2. Repeated L-LTP switches $W_2$ to the high state and elevates synaptic $Ca^{2+}$ (red trace, time-averaged and scaled as in Fig. 5A) and the tag. Arrows, imposed decreases of activity at synapse 2. Between state transitions of $W_2$, smaller fluctuations of synapse 2 variables occur as a consequence of switches in the states of synapses 1, 3, and 4.

## TABLE 1

## Standard Parameter Values for the Reduced Model

| Parameters and Values | Biochemical Significance |
|---|---|
| $Ca_{tet}$ = 0.9 µM, $Ca_{bas}$ = 0.04 µM, $K_{max}$ = 0.7, $Ca_{limit}$ = 1.2 µM, $f_{nuc}$ = 0.25, $K_{sum}$ = 0.02 µM | Parameters describing dynamics of synaptic and nuclear $[Ca^{2+}]$ ($Ca_{syn}$, $Ca_{nuc}$) during imposed stimuli and during ongoing neuronal activity. |
| $k_{phos}$ = 13,300 hr$^{-1}$, $k_{deph}$ = 0.33 hr$^{-1}$, $k_{syn}$ = 4,170 hr$^{-1}$, $k_{deg}$ = 0.33 hr$^{-1}$, $K_{SP}$ = 0.6 µM, $K_{ST}$ = 0.7 µM | Parameters describing the dynamics of TAG, GPROD, and the sigmoid activation functions $S_P$ and $S_T$. |
| $k_{ltp}$ = 0.4 hr$^{-1}$, $k_{ltpbas}$ = 0.001 hr$^{-1}$, $k_{ltdbas}$ = 0.01 hr$^{-1}$, $W_{max}$ = 1.15, $W_{inc}$ = 0.15 | Parameters describing changes in W during long-term potentiation and passive decay, and synaptic scaling parameters ($W_{max}$ and $W_{inc}$, used in Fig. 8). |

# Figure 1

A

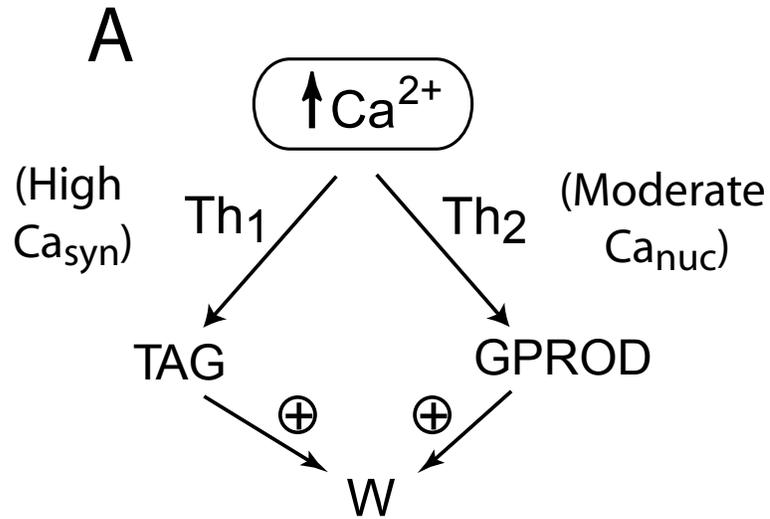

B **Synapses**

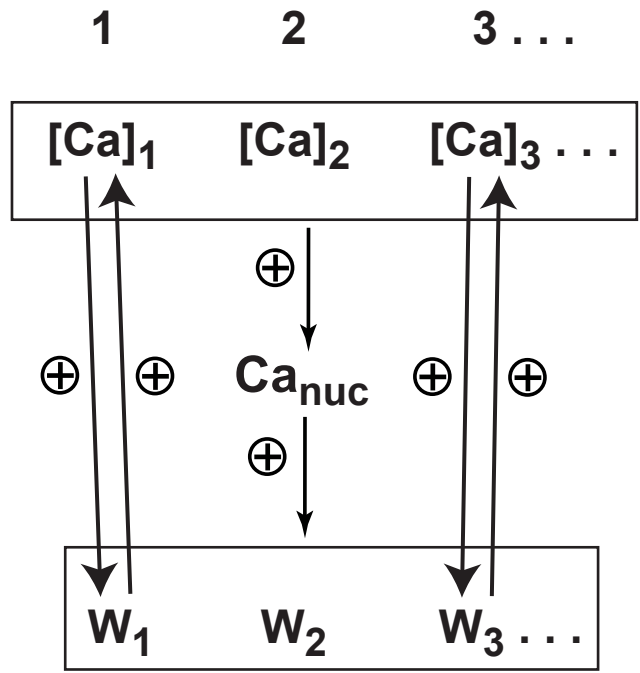

# Figure 2

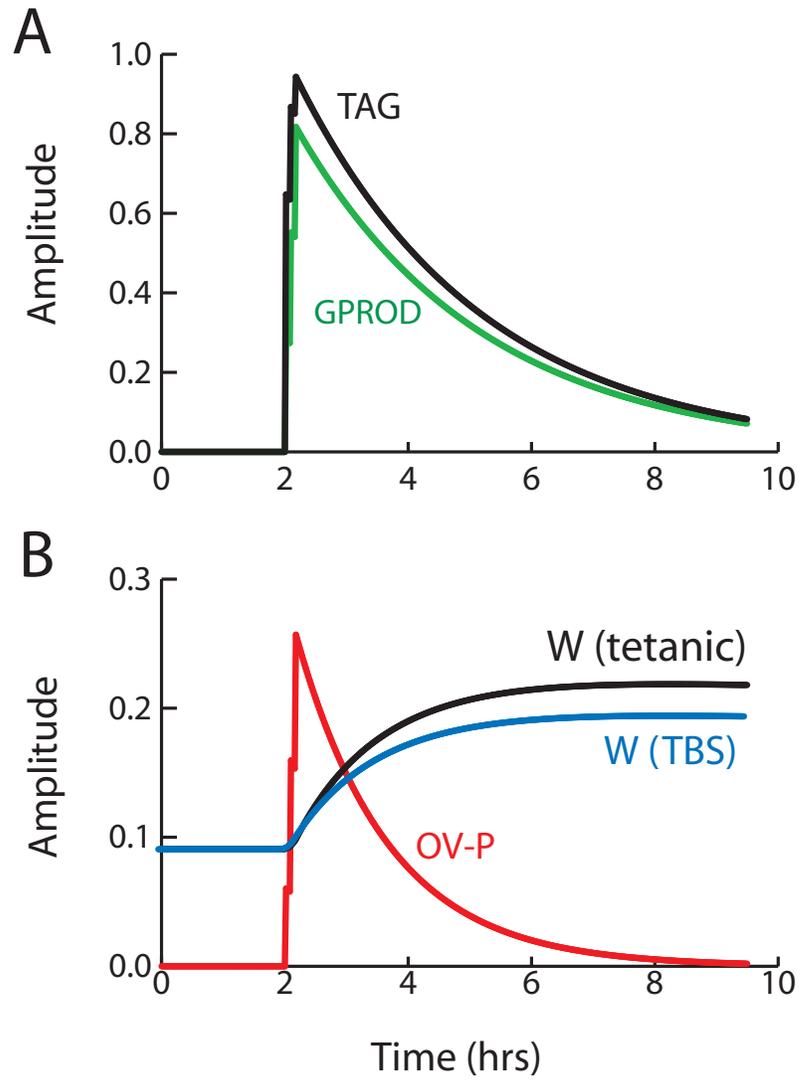

# Figure 3

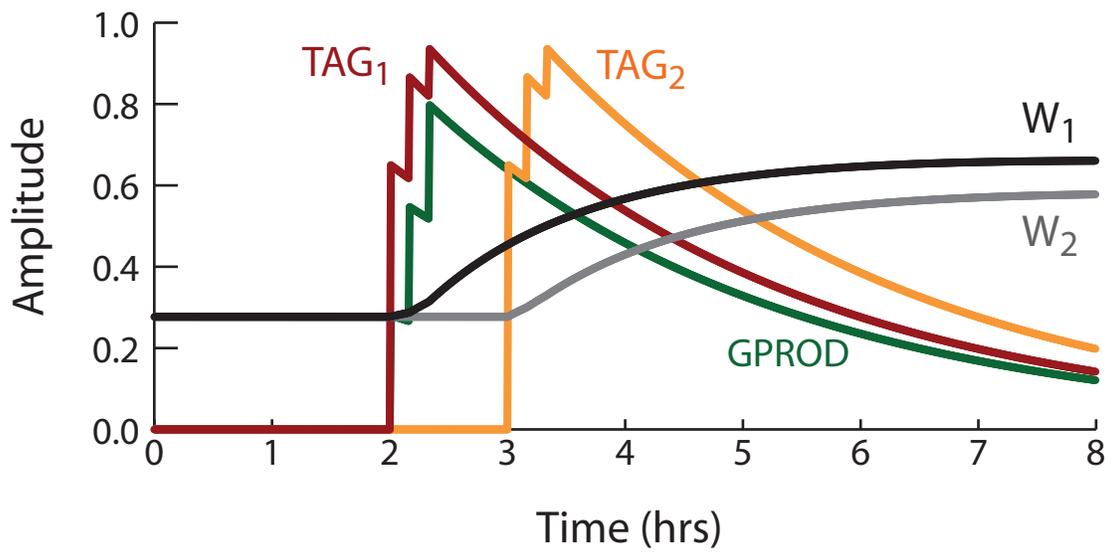

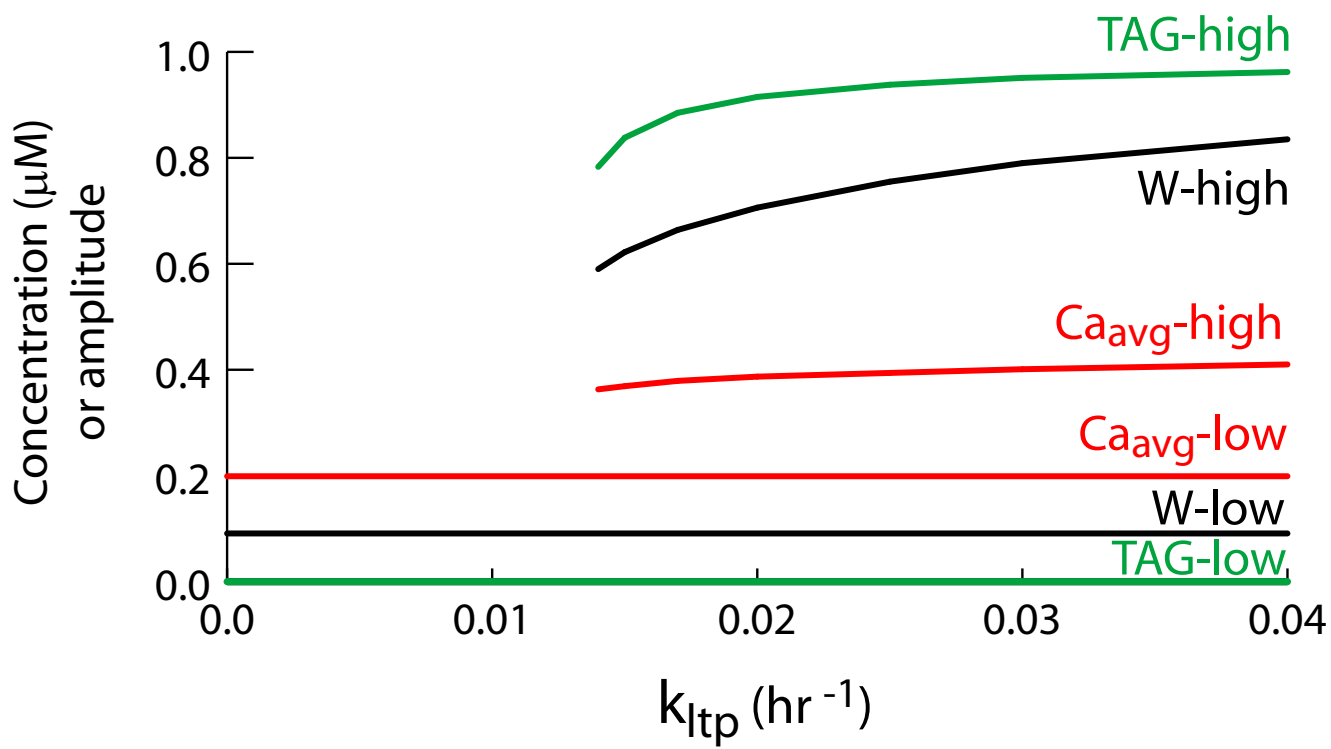

**Figure 5**

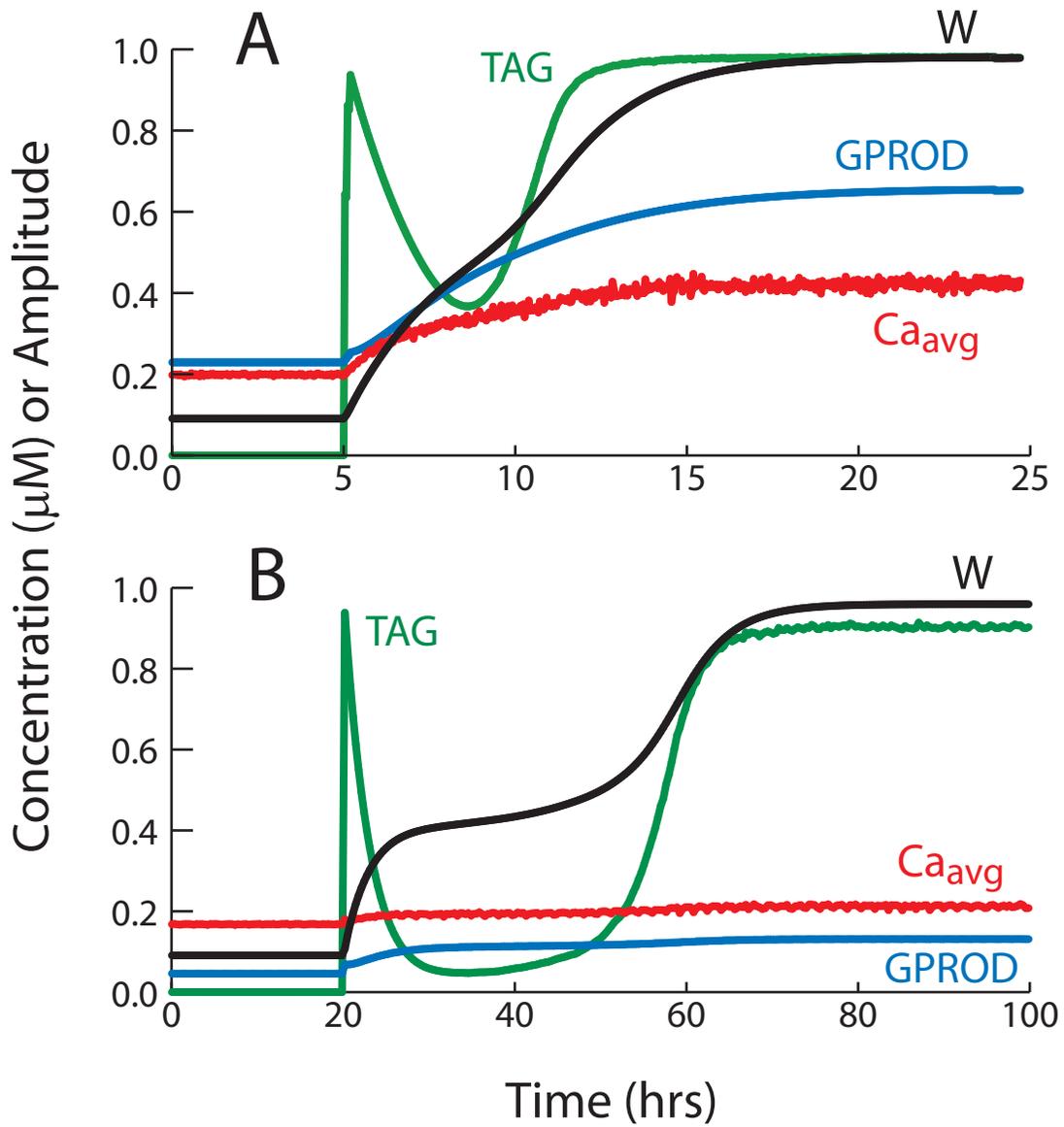

# Figure 6

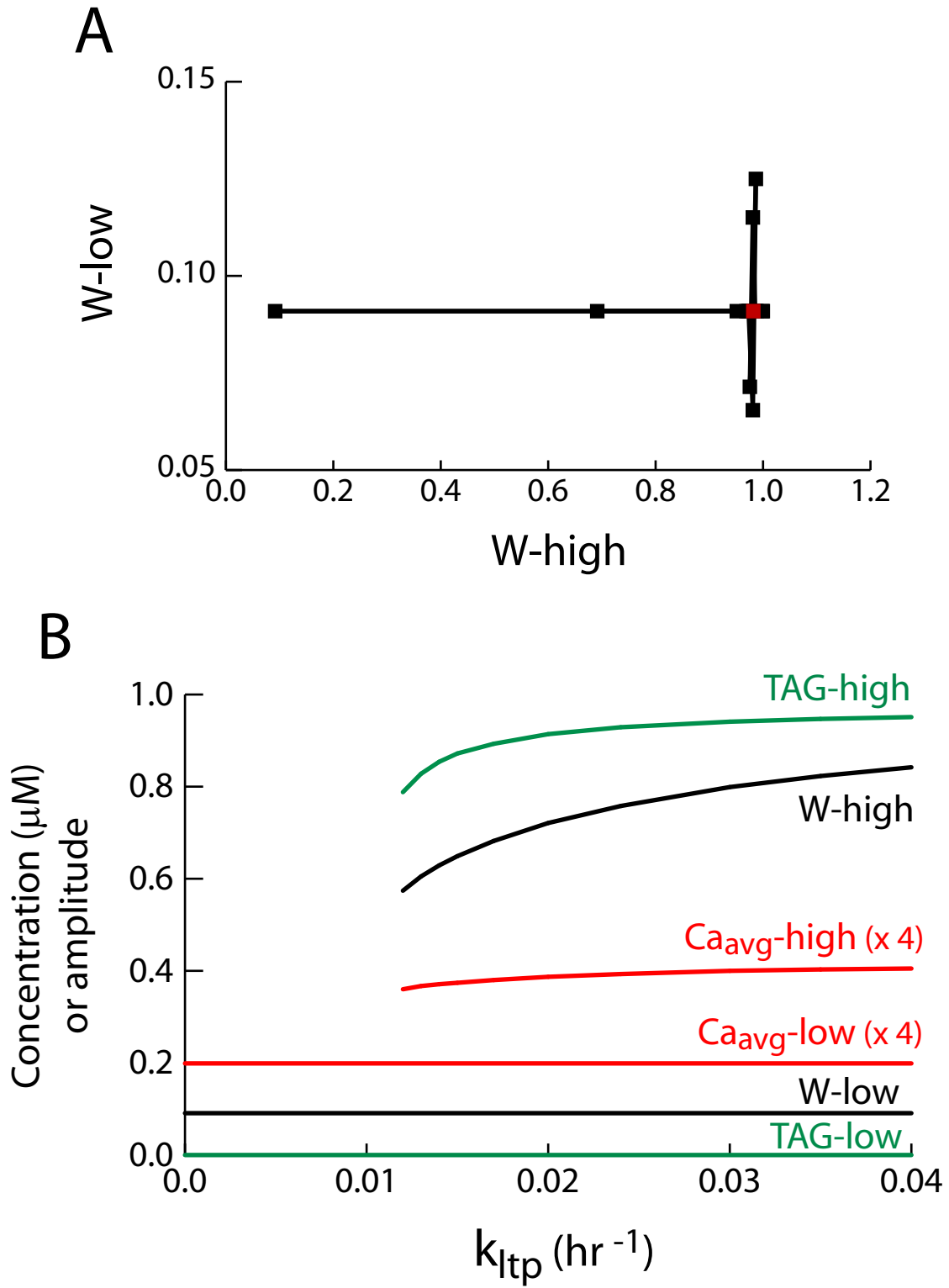

# Figure 7

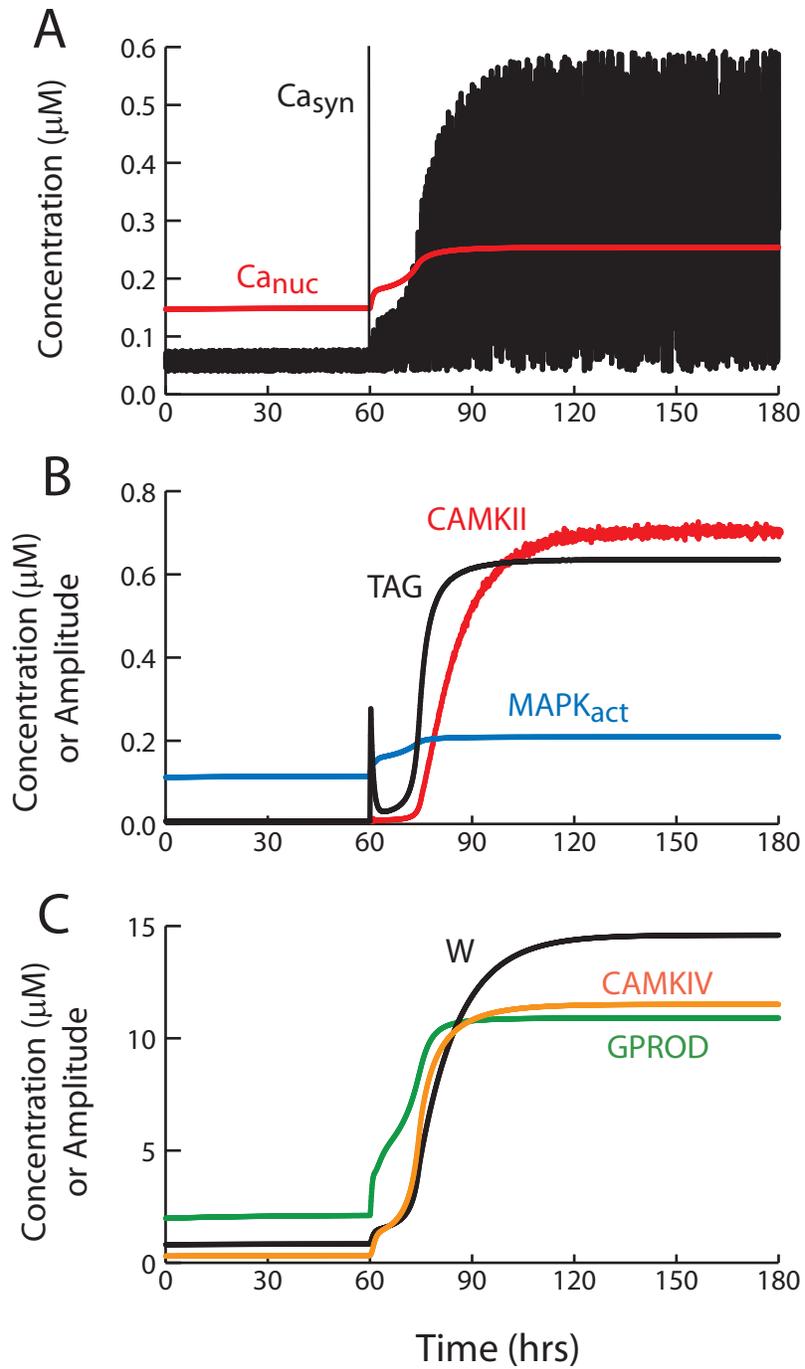

# Figure 8

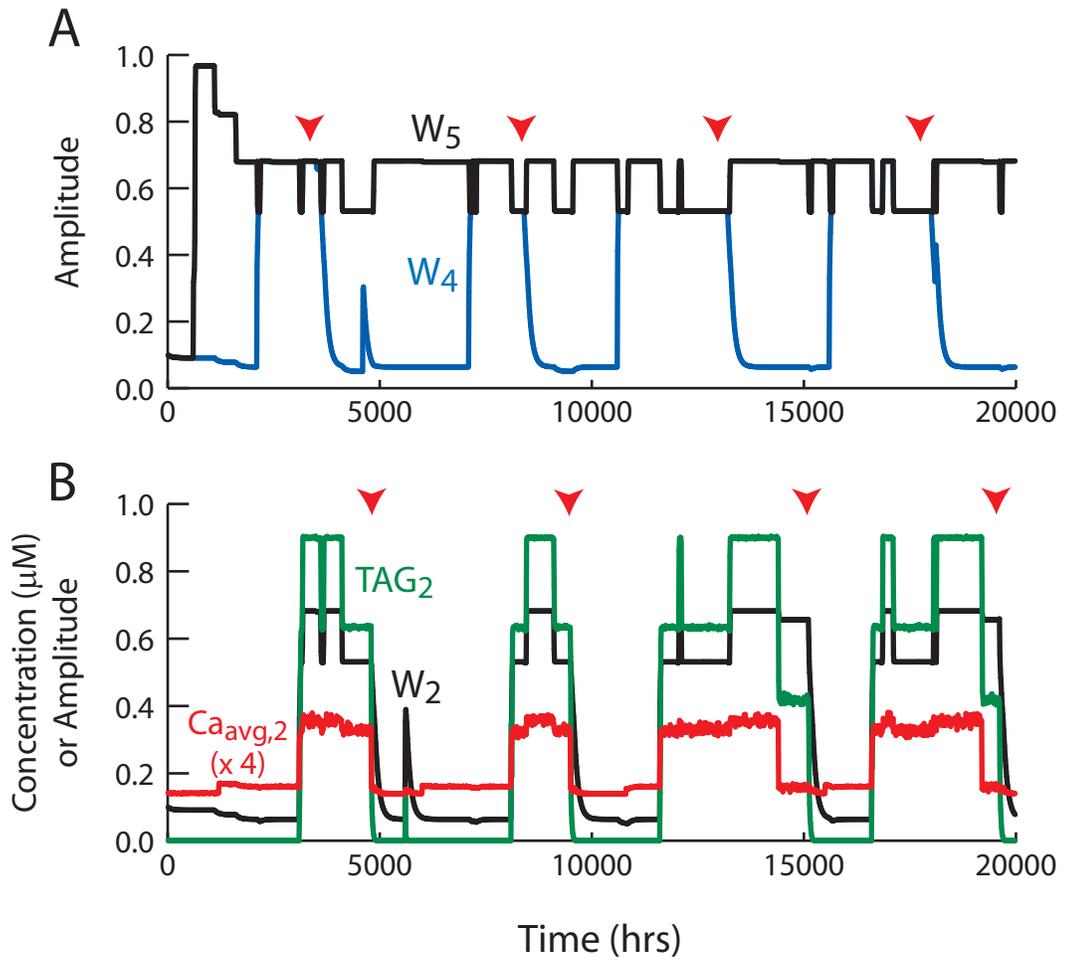